# Spincaloritronic signal generation in non-degenerate Si

Naoto Yamashita [1,#], Yuichiro Ando [1,#], Hayato Koike [2],
Shinji Miwa [3], Yoshishige Suzuki [3] and Masashi Shiraishi [1,#,$]

1. Department of Electronic Science and Engineering, Kyoto University, Kyoto 615-8510, Japan.
2. Technology HQ, TDK Corporation, Chiba 272-8558, Japan.
3. Graduate School of Engineering Science, Osaka University, Toyonaka 560-8531, Japan.

\# These authors equally contributed to this work.
$ Corresponding author (mshiraishi@kuee.kyoto-u.ac.jp)

**Spincaloritronic signal generation due to thermal spin injection and spin transport is demonstrated in a non-degenerate Si spin valve. The spin-dependent Seebeck effect is used for the spincaloritronic signal generation, and the thermal gradient of about 200 mK at an interface of Fe and Si enables generating a spin voltage of 8 $\mu$V at room temperature. A simple expansion of a conventional spin drift-diffusion model with taking into account the spin-dependent Seebeck effect shows semiconductor materials are quite potential for the spincaloritronic signal generation comparing with metallic materials, which can allow efficient heat recycling in semiconductor spin devices.**

Spincaloritronics, which combines spintronics and thermoelectricity [1], is a new field of spintronics and attracts strong attention. In spintronics, spin currents, such as pure spin currents and spin-polarized currents, are used to propagate information. Whereas spin currents have been generated electrically and dynamically in a wide variety of materials [2-10], the utilization of heat currents to generate spin currents is a newly established method, and the method spawned the field of spincaloritronics [11-17]. The coupling of spin current and heat current leads novel physical phenomena, with a number of new attractive caloritronic effects such as the spin-Seebeck effect [13,14], the spin-dependent Seebeck effect [15], and the spin Peltier effect [11,16]. However, materials used to generate these spincaloritronic effects have been limited to metals [15,16,18,19] and magnetic insulators [14] and the methodology has not yet been widely extended to semiconductors. One example claiming detection of a spincaloritronic effect in semiconductor is the Seebeck spin tunneling [20], where a controversial 3-terminal method (only one ferromagnetic electrode is used to observe the Hanle-like signals) [21] was used and thermal spin accumulation (not spin transport) was claimed. However, the reliability of the 3-terminal method has been under strong debate [22-29]. Thus, it is noteworthy to realize spin transport in semiconductors by using spincaloritronic effects. Because the spin-Seebeck effect expresses only in a ferrimagnetic insulator, the spin-dependent Seebeck effect is a potential candidate to realize thermally induced spin injection and transport in semiconductors.

In this paper, we report spincaloritronic signal generation in non-degenerate Si, the most common material in semiconductor electronics, by utilizing the spin-dependent Seebeck effect. The thermal gradient, generated by electric current injection, at an interface between a ferromagnetic electrode and a non-degenerate Si spin channel enables generation of a spin current in the Si. A simple expansion of a conventional spin drift-diffusion model with taking into account the spin-dependent Seebeck contribution shows that a semiconductor is more potential to generate larger thermal spin signals under the same thermal gradient comparing with a metal.

A non-degenerate Phosphorus-doped ($n \approx 2 \times 10^{18}$ cm$^{-3}$) silicon spin valve was fabricated on a silicon-on-insulator (SOI) substrate with a structure of Si (100 nm)/SiO$_2$ (200 nm)/bulk Si (see Fig. 1(a)), which is the same structure of spin MOSFETs. The conductivity of the Si channel was measured by using a four-probe method. The upper Si layer was phosphorous (P)-doped by ion implantation. Ferromagnetic tunnel junctions were formed on silicon with higher doping level (n$^+$-Si; $t$=20 nm; $n \approx 5 \times 10^{19}$ cm$^{-3}$) using an MgO tunnel barrier (0.8 nm thick) and 17-nm-thick Fe thin film. After the natural oxide layer on the Si channel was removed using an HF solution, Pd(3 nm)/Fe(13 nm)/MgO(0.8 nm) was grown on the etched surface by molecular beam epitaxy. Then, we etched out the Pd (3 nm)/Fe (3 nm) layers, and Ta (3nm) was grown on the remaining Fe. The contacts had dimensions of 0.5×21 μm$^2$ and 2×21 μm$^2$, respectively. The Si channel surface and sidewalls at the ferromagnetic contacts were buried by SiO$_2$. The nonmagnetic electrodes, with dimensions of 21×21 μm$^2$, were made from Al and produced by ion milling. The center-to-center distance between the FM electrodes, $L$, was set to be 1.75 μm. The spin valve characteristics were investigated by using a probing station (Janis Research Company Inc., ST-500), a source meter (Keithley Instruments, 2400 and 2401), and a digital multimeter (Keithley Instruments, 2010). The measurements for detecting spincaloritronic signals were implemented using an AC lock-in technique, where the AC frequency was set to 17 Hz applied using LI5655 (NF Corporation). All measurements were performed at RT.

For the measurements for detecting spincaloritronic signals in Si, we used an electrical local three-terminal magnetoresistance (3T-MR) method [30,31]. To note is that this 3T-MR is a different method from the controversial 3-terminal method [21]. We use two ferromagnetic electrodes and measure magnetoresistance in addition to the Hanle signals, which provides strong evidence of successful spin injection and transport in Si [30,31]. The other advantage of this method is to obtain larger spin signals due to spin drift comparing with a non-local 4-terminal method, where only spin diffusion contributes to spin transport [31].

Because the spincaloritronic signal due to the spin-dependent Seebeck effect scales with the Joule heating, we used an AC lock-in technique to extract the thermally induced spincaloritronic signals, as described in ref. [15]. The output voltage, $V_{detect}$, as a function of the injected current $I$ is described by: $V_{detect} = R_1 I + R_2 I^2 + \cdots$, with $R_1$ and $R_2$ are parameters relevant to electrically-generated spin signal and spin caloritronics signal, respectively. We measured the 2$^{nd}$ harmonic term, $R_2 I^2$, linked to spin caloritronics signal using the AC lock-in technique, whereas the spin signal coming from electrical spin injection is measured via DC detection. Some issues can hinder precise estimation of spincaloritronic signal components using the AC lock-in measurement scheme. Indeed, the MgO tunneling barrier induces nonlinearity in $I$-$V$ characteristics, which generates spurious components in the 2$^{nd}$ harmonic signals. In addition, nonlinear bias dependence of spin signals [32] also contributes to the generation of spurious 2$^{nd}$ harmonic signals. However, a particular attention, described below, was paid to eliminate these spurious effects in our study.

Figures 2(a) and 2(b) show the measurement setup and the thermally induced magnetoresistance observed in a non-degenerate Si, respectively. The DC and AC bias voltages were set to be 3.0 V and 1.0 V root-mean-square (rms), respectively. The AC frequency was set to be 17 Hz to avoid unnecessary thermal contribution to the experiment [15]. As can be seen in Fig. 2(b), a 2$^{nd}$ harmonic voltage, $R_2I^2$, of 19 μV was measured, including the spincaloritronic signal and also spurious signals, described in the previous paragraph. From the Fourier transformation analysis using the *I-V* curve and the bias dependence of spin signals to estimate these spurious contributions, which is described in more detail in the discussion part, the contribution from the spurious effects is estimated to be 11 μV for an AC injection bias of 1.0 V. Figure 2(c) shows the spincaloritronic signals as a function of the square of the AC electric current after removing spurious signals. The experimental results exhibit clear $I^2$ dependence, providing evidence that the 2$^{nd}$ harmonic voltage can be attributed to the spincaloritronic effect.

In order to estimate the thermal gradient in our device, a model was constructed by combining the conventional spin-drift-diffusion model and the spin-dependent Seebeck effects. Since a bias electric field is applied to the Si spin channel, the upstream and downstream spin transport length scales ($\lambda_u$ and $\lambda_d$) were used instead of the spin diffusion length of Si ($\lambda_N$), where $\lambda_{u(d)} = [+(-)\frac{|eE|}{2k_BT} + \sqrt{(\frac{eE}{2k_BT})^2 + (\frac{1}{\lambda_N})^2}]^{-1}$ (*E* is the electric field in the spin channel and $k_B$ is the Boltzmann constant) [33]. Furthermore, terms due to the spin-dependent Seebeck effect are included in the description of up- and down-spin currents as, $j_s = -\sigma_s(\frac{\partial V_s}{\partial z} + S_s \nabla T)$, ($s = \uparrow, \downarrow$), where $j_s$ is the spin current, $\sigma_s$ is the conductivity, *z* is a position coordinate, $V_s$ is the spin-dependent voltage ($V_s = \mu_s/e$, $e(<0)$ is the electric charge and $\mu_s$ is the electrochemical potential), $S_\uparrow$ and $S_\downarrow$ are the Seebeck coefficient of up and down spins of Fe, respectively, and $T$ is the temperature [15]. We took into account for this model: (i) the device structure with a contribution from the spin-dependent interface tunnel resistance due to the MgO (the thickness of the MgO is neglected), (ii) the back flow of the spin current from FM2, and (iii) the spin drift effect from electrical current injection into Si (see also ref. [31]). The detail of the theoretical model is shown in Sec. A of Supplemental Materials [34]. Experimentally, we measured the conductivities of the Fe and the Si at RT to be $\sigma_F = 8.3 \times 10^6$ Sm$^{-1}$ and $\sigma_N = 2.12 \times 10^3$ Sm$^{-1}$, also the interface RA was measured to be $2.56 \times 10^{-9}$ $\Omega$m$^2$ (the spin injection side: FM2) and $4.66 \times 10^{-9}$ $\Omega$m$^2$ (the spin extraction side: FM1). During the measurement, the electric field in the Si spin channel was set to $1.71 \times 10^6$ V/m, $L$ was 1.75 μm and $\lambda_N$ was 5.0 μm (from electrical spin injection experiments. See Sec. B. of Supplemental Materials [34]). Using values from literatures, we set $\alpha_{F1}(=\alpha_{F3})$ = 0.4 [35], $S$ = 15 μV/K at 293 K [36] and $\lambda_F$ = 9 nm [37] for Fe at RT. The spin-dependent Seebeck coefficient defined as $S_S = S_\uparrow - S_\downarrow$ is theoretically described as $S_S = PS$ [15,38], and the value for Fe was calculated to be 6 μV/K. Consequently, the temperature gradient between the Fe and the Si was estimated to be about 200 mK for a spincaloritronic signal of 8 μV.

A particular attention was given to eliminate spurious signals from our estimation of thermal spin signal magnitude in silicon. These spurious contributions are due to the nonlinearity of *I-V* curves and the bias dependence of the spin signals. In this measurement scheme, rms AC voltage of 1.0 V was applied to the device in addition to a constant bias voltage (3.0 V), and the nonlinearities around the dc bias voltage, 3.0 V, are superposed on the spin caloritronic signals. We estimated their influence by deducing, from experimental data, expecting signal of spurious contribution at 17 Hz. For estimating spurious signal at 17 Hz, first we measured the electric current as a function of the dc voltage applied in our 3T-MR setup and fitted the results by using a fifth-order polynomial function: $I=G_1V_{inj}+G_2V_{inj}^2+...+G_5V_{inj}^5$ (shown as dashed lines in Fig. 3(a)), where $G_i$ ($i=1\sim5$) is the $i$-th order conductance and $V_{inj}$ is the injection voltage. In the same setup, we also measured the electrical spin signal in a dc configuration as a function of the applied electric current. We associated this spin signal, noted as $V_{DC\ spin\ signal}$, to electrical spin injection and spurious effects. We fitted the results using also fifth-order polynomial function: $V_{DC\ spin\ signal} = Rs_1I+Rs_2I^2+...+Rs_5I^5$ by substituting the previous fitted function for $I$: $V_{DC\ spin\ signal} = \sum_{i=1}^{5}R_{s,i}(\sum_{k=1}^{5}G_k(V_{inj})^k)$, where $R_i$ ($i=1\sim5$) is the $i$-th order resistance (see Fig. 3(b)). By determining the coefficients, we are now able to write the time-dependent contribution as $V_{DC\ spin\ signal}(t)=\sum_{i=1}^{5}R_{s,i}(\sum_{k=1}^{5}G_k(V_{inj}(t)))$ under AC+DC excitation because in this case $V_{inj}(t) = V_{dc} + V_0\sin(2\pi ft)$, where $f$ is AC frequency (17 Hz), $V_{DC}$ is 3.0 V and $V_0$ is 1.0 V in this experiment. Using the Fourier transformation, the 2$f$ component of the $V_{DC\ spin\ signal}(t)$ was calculated (see also Sec. C of Supplemental Materials [34]). Figure 3(c) shows the comparison between the expected 2[nd] harmonic components from the spurious effects and the experimental signal obtained using the lock-in method. A clear difference appears between the two signals: for example at an rms AC voltage of 1.0V the difference is 8 μV. Finally, the spincaloritronic signals are obtained by subtracting the spurious signals from the experimentally measured 2[nd] harmonic signals, and they exhibit quadratic dependence as a function of the rms AC voltage (see Fig. 3(d)), which provides an evidence of successful spin caloritronics signal detection from a non-degenerate Si spin valve.

Figures 4(a) and 4(b) show contour plots of the magnitude of spin caloritronic signals as a function of conductivity and spin diffusion length, and of spin polarization and spin-dependent Seebeck coefficients, respectively. In the calculations, the other parameters, such as the thermal gradient, are set to be the same as those in estimating the thermal gradient of our Si spin devices. Figure 4(a) shows an enhancement of spincaloritronics signal by increasing the spin polarization or spin-dependent Seebeck coefficient of a ferromagnetic electrode. Thus, an introduction of Heusler-based half metal, such as CoFeAl [18], with high spin polarization and large spin-dependent Seebeck coefficient could greatly enhance spincaloritronics signals. Furthermore, long spin diffusion length allows efficient generation of spincaloritronic signals (see Fig. 4(b)), because in the spin-dependent Seebeck effect, spin current generated by heat propagates in a nonmagnetic material and is detected by a detector ferromagnet apart from a injector ferromagnet. Regarding the conductivity, there is an optimum value for the spin channel conductivity, which is understood as manifestation of the conductance mismatch problem [39] as in the case of electrical spin injection and transport in semiconductors. It is notable that a material with metal-like conductivity ($\sim 10^6$ Sm$^{-1}$) cannot generate larger spincaloritronic signal compared with a material with lower conductivity ($\sim 10^3$ Sm$^{-1}$). In fact, semiconductors can generate about a 3-orders of magnitude larger spincaloritronic signal in this calculation, which is a significant advantage of semiconductors.

Difference in physical nature between the spincaloritronic effect in our study and the other spin-related caloritronics effects is discussed in this paragraph. As describe above, we used spin-dependent Seebeck effect [15], which is the spin-version of the conventional charge Seebeck effect. The nature of the spin Seebeck effect [13,14], the other significant spincaloritronic effect, is due to temperature difference between magnon and electron systems in a ferrimagnetic insulator and a heavy metal. In the magneto-Seebeck effect [40], the origin of the effect is difference of the charge Seebeck coefficients under parallel and antiparallel magnetization configurations in magnetic tunnel junctions and the magneto-Seebeck signal appears in a DC component of magnetoresistance. Thus, the origins of the spin Seebeck and magneto-Seebeck effects are different from that in the spin-dependent Seebeck effect, so these effects can be ruled out in our study. The "Seebeck spin tunneling" can be another effect appearing in Si [20], and the electrical 3-terminal method [21] (not the same method we used in this study), which is under strong debate [22-29], was used. The detail of essential difference between the results in ref. [20] and in our study is discussed in Sec. D of Supplemental Materials [34], and we simply emphasize that we for the first time demonstrated the spin injection and transport in a semiconductor channel (Si) using a spincaloritronic effect.

In summary, we achieved spincaloritronic signal generation in non-degenerate Si, utilizing the spin-dependent Seebeck effect. Thermal gradient at the interface between Fe and Si allows the generation of spin current in the Si, which was detected as the $2^{nd}$ harmonic component of the spin accumulation voltages at the detector ferromagnet. A simple expansion of the conventional spin drift-diffusion model with taking into account the spin-dependent Seebeck contribution reproduces the experimental result, and also indicates that semiconductor materials are more potential in the heat recycling comparing with metallic materials. Our study also shows that the approach in this study is applicable to heat recycling in Si-based devices, such as spin MOSFETs.

**Acknowledgement**

A part of the research was supported in part by a Gran-in-Aid for Scientific Research from the Ministry of Education, Culture, Sports, Science and Technology (MEXT) of Japan, Scientific Research (S) "Semiconductor Spincurrentronics" (No. 16H0633). M.S. thanks Dr. F. Rortais for his critical reading of the manuscript.

**Figures and figure captions**

**Figure 1 (a)** Non-degenerate ($n \approx 2 \times 10^{18}$ cm$^{-3}$) Si spin valves were fabricated on a silicon-on-insulator (SOI) substrate. The ferromagnetic tunnel junctions were formed on n$^+$-Si (20 nm thick with a doping concentration of $5 \times 10^{19}$ cm$^{-3}$) using an MgO tunnel barrier (0.8 nm thick) and 17-nm-thick Fe thin film. After nanofabrication of the two ferromagnetic electrodes, the center-to-center length, $L$, was 1.75 μm. **(b)** A schematic illustration of our device to calculate the magnitude of spincaloritronic signals. In this model, the device is divided in five regions: the red and black regions are ferromagnetic (FM) and nonmagnetic (NM), respectively. An electric current flows from FM1 to FM2 i.e. spins are injected from FM2. We set an open circuit condition for the calculation. For more detail, see Sec. A of Supplemental Materials [34].

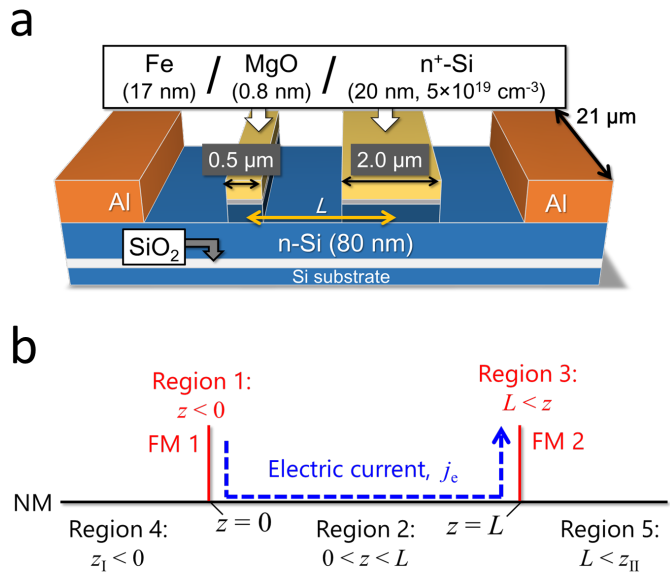

Fig. 1    N. Yamashita et al.

**Figure 2 (a)** A scheme of measuring spincaloritronic signals from the non-degenerate Si spin valves. The ac lock-in technique was used to detect the 2$^{nd}$ harmonic spin signals attributed to the spincaloritronic effect. **(b)** A 2$f$ spin signal observed from the Si spin valve at RT. The external magnetic field was swept from negative to positive (the red solid line) and from positive to negative (the blue solid line), and a 2$^{nd}$ harmonic voltage of 19 µV was observed. Note that spurious signals in addition to the spincaloritronic signal are included in this 2$f$ signal. **(c)** $I_{AC}^2$ dependence of the spincaloritronic signals. The detail of separating the spincaloritronic signals and spurious signals is described in the main text. The black closed squares are experimental results and the red solid line shows the fitting line. The experimental results were well fit.

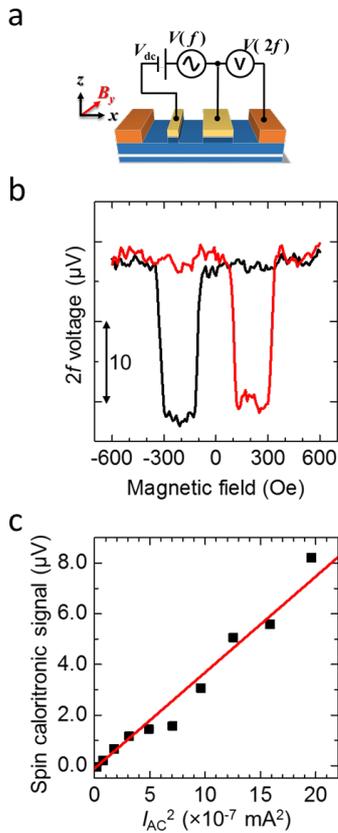

Fig. 2  N. Yamashita et al.

**Figure 3 (a)** The bias voltage dependence of the electric current in the non-degenerate Si spin valve at around 3.0 V, the dc offset voltage in the experiment. The black closed squares are experimental results and the red solid line is the fitting curve (the 5$^{th}$ order polynomial function). **(b)** The electric current dependence of the electrical spin signal. The black closed squares are experimental results and the red solid line is the fitting curve (the 5$^{th}$ order polynomial function). **(c)** Comparison of the measured 2*f* spin signal and the spurious signal. Calculation of magnitude of the spurious signals is described in the main text. The spincaloritronic signal is obtained by subtracting the spurious signals from the measured 2*f* spin signal. **(d)** The spincaloritronic signal as a function of the applied ac voltage with dc offset voltage of 3.0 V. The red solid line is the quadratic fitting function.

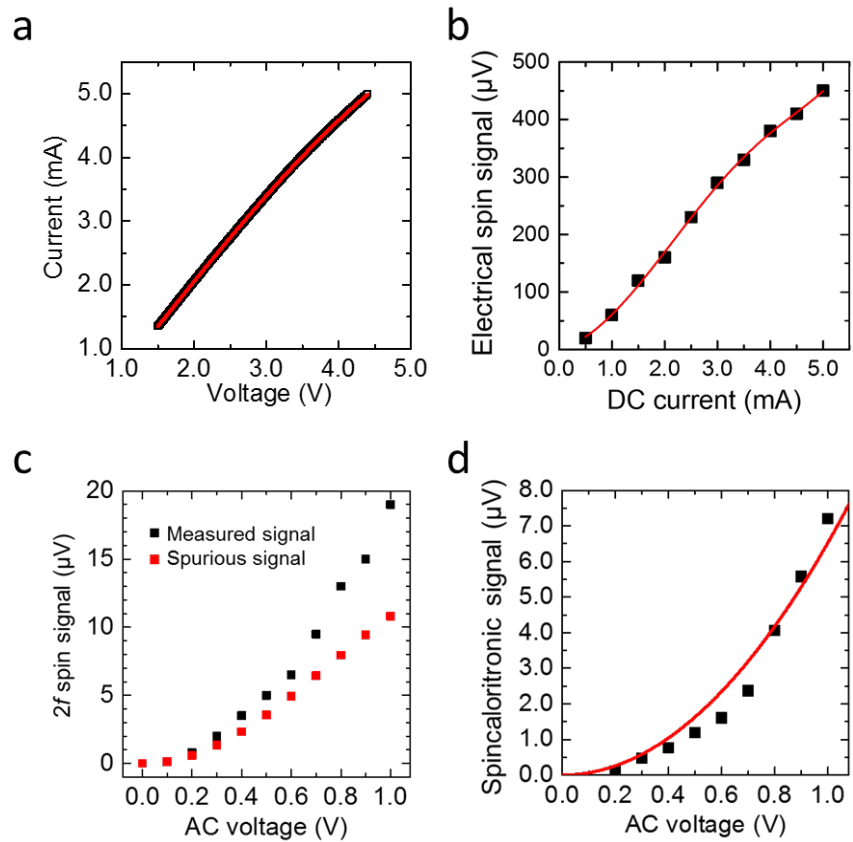

**Fig. 3**   N. Yamashita et al.

**Figure 4 (a)** A contour plot of spincaloritronic signals by changing spin polarization and spin-dependent Seebeck coefficient of ferromagnet. **(b)** A contour plot of magnitude of spincaloritronic signals by changing spin diffusion length and conductivity of a spin channel.

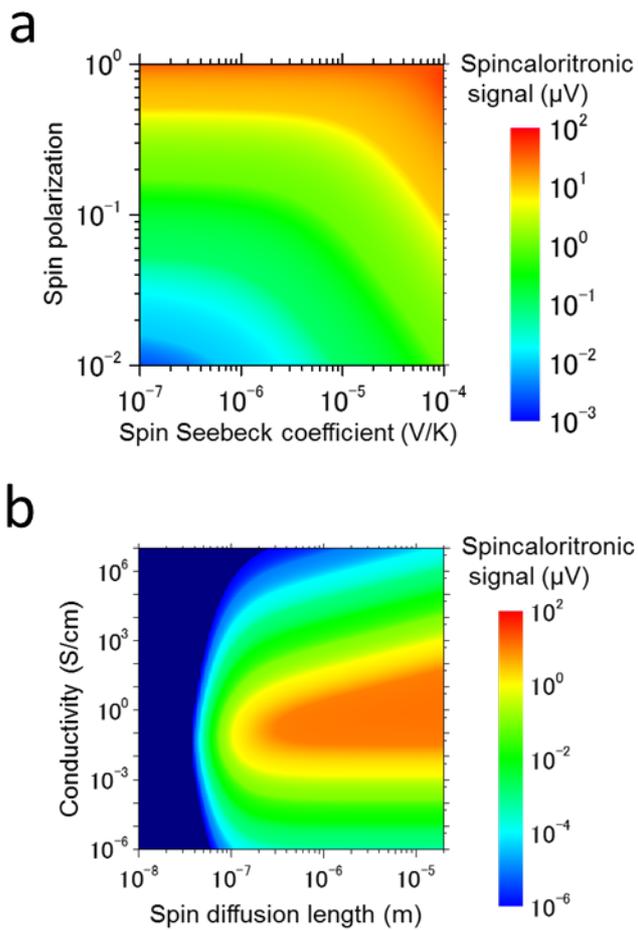

**Fig. 4   N. Yamashita et al.**

*Supplemental Materials*

## Spincaloritronic signal generation in non-degenerate Si

Naoto Yamashita [1,#], Yuichiro Ando [1,#], Hayato Koike [2],
Shinji Miwa [3], Yoshishige Suzuki [3] and Masashi Shiraishi [1,#,$]

1. Department of Electronic Science and Engineering, Kyoto University, Kyoto 615-8510, Japan.
2. Technology HQ, TDK Corporation, Chiba 272-8558, Japan.
3. Graduate School of Engineering Science, Osaka University, Toyonaka 560-8531, Japan.

**A. Theoretical model building**

By considering an electric current flowing from FM1 to FM2, i.e. spins (electrons) are injected from FM2, the spin-dependent voltages and spin current densities in FM1 (region 1) and FM2 (region 3) can be written as :

$$V_{1,\uparrow(\downarrow)} = -\frac{j_e}{\sigma_F}z + (-)V_1^- \frac{\sigma_{\downarrow(\uparrow)1}}{\sigma_F}\exp(\frac{z}{\lambda_F}) + D,$$

$$j_{1,\uparrow(\downarrow)} = \frac{\sigma_{\uparrow(\downarrow)1}}{\sigma_F}j_e - (+)\frac{V_1^-}{R_F}\exp(\frac{z}{\lambda_F}) - \sigma_{\uparrow(\downarrow)1}S_{\uparrow(\downarrow)1}\nabla T_0,$$

(FM, Region 1) (1)

$$V_{3,\uparrow(\downarrow)} = -\frac{j_e}{\sigma_F}(z-L) - \frac{j_e}{\sigma_N}L + (-)V_3^+ \frac{\sigma_{\downarrow(\uparrow)3}}{\sigma_F}\exp(-\frac{z-L}{\lambda_F}) - E,$$

$$j_{3,\uparrow(\downarrow)} = \frac{\sigma_{\downarrow(\uparrow)3}}{\sigma_F}j_e + (-)\frac{V_3^+}{R_F}\exp(-\frac{z_{II}-L}{\lambda_F}) - \sigma_{\uparrow(\downarrow)3}S_{\uparrow(\downarrow)3}\nabla T_L,$$

(FM, Region 3) (2)

where, $j_e$, $\lambda_F$, $R_F$, $\sigma_{Fi}$, $D$, $E$, and $L$ are the electric current, the spin diffusion length of FM, the spin resistance of FM (defined as $R_F = \lambda_F(\sigma_\uparrow^{-1} + \sigma_\downarrow^{-1})$), the conductivity of FM "$i$( = 1, 3)" (= $\sigma_{\uparrow i} + \sigma_{\downarrow i}$), the spin accumulation voltage at the interface between FM1 and NM, that between FM2 and NM, and the gap length between FM1 and FM2, respectively (see also Fig. 1(b) in the main text). $\nabla T_0$ and $\nabla T_L$ are the temperature gradients at the point of $z = 0$ and $z = L$. $V_1^-$ and $V_3^+$ are the spin-dependent voltages at FM1 and FM2. Similarly, the spin-dependent voltages and the spin current densities in the three NM regions (regions 2, 4 and 5), including the spin drift effect, are written as:

$$V_{2,\uparrow(\downarrow)} = +(-)\frac{V_2^+}{2}\exp(-\frac{z}{\lambda_u}) + (-)\frac{V_2^-}{2}\exp(\frac{z-L}{\lambda_d}) - \frac{j_e}{\sigma_N}z,$$

$$j_{2,\uparrow(\downarrow)} = +(-)\frac{V_2^+}{R_u}\exp(-\frac{z}{\lambda_u}) - (+)\frac{V_2^-}{R_d}\exp(\frac{z-L}{\lambda_d}) + \frac{j_e}{2} - \frac{1}{2}\sigma_N S_N \nabla T_N,$$

(NM, Region 2)  (3)

$$V_{4,\uparrow(\downarrow)} = +(-)\frac{V_4^-}{2}\exp(\frac{z_I}{\lambda_N}),$$

$$j_{4,\uparrow(\downarrow)} = -(+)\frac{V_4^-}{R_N}\exp(\frac{z_I}{\lambda_d}) - \frac{1}{2}\sigma_N S_N \nabla T_N,$$

(NM, Region 4)  (4)

$$V_{5,\uparrow(\downarrow)} = -\frac{j_e}{\sigma_N}L + (-)\frac{V_5^+}{2}\exp(-\frac{z_{II}-L}{\lambda_N}),$$

$$j_{5,\uparrow(\downarrow)} = +(-)\frac{V_5^+}{R_N}\exp(-\frac{z_{II}-L}{\lambda_N}) - \frac{1}{2}\sigma_N S_N \nabla T_N,$$

(NM, Region 5)  (5)

$$R_{N(d,u)} = \frac{4}{\sigma_N}\lambda_{N(d,u)}, \quad (6)$$

where $\nabla T_N$ is the thermal gradient in NM, and $\sigma_0 (= 2\sigma_N)$ is the conductivity of NM. The continuity conditions at $z = 0$ and $L$ for spin voltages and spin currents for up and down spins, including the interfacial spin-dependent tunneling resistance, are set to be

$$V_{1,\uparrow(\downarrow)} - R_{i1\uparrow(\downarrow)}j_{1\uparrow(\downarrow)} = V_{2,\uparrow(\downarrow)} = V_{3,\uparrow(\downarrow)},$$
$$j_{1,\uparrow(\downarrow)} = j_{2,\uparrow(\downarrow)} + j_{3,\uparrow(\downarrow)},$$
(at $z=0$)  (7)

$$V_{5,\uparrow(\downarrow)} + R_{i3\uparrow(\downarrow)}j_{1\uparrow(\downarrow)} = V_{3,\uparrow(\downarrow)} = V_{4,\uparrow(\downarrow)},$$
$$j_{3,\uparrow(\downarrow)} = j_{4,\uparrow(\downarrow)} + j_{5,\uparrow(\downarrow)},$$
(at $z=L$)  (8)

where $R_{i1\uparrow(\downarrow)}$ and $R_{i3\uparrow(\downarrow)}$ are the spin-dependent resistances due to the tunneling barrier at the interface between FM1 and NM and between FM2 and NM, respectively. The spin accumulation voltage probed in the 3T-MR scheme, written $D$ in Eq. (1), can be described under the parallel magnetization configuration as:

$$D^{parallel} = \{\frac{1}{2}\alpha_{F1} + (\sigma_{\uparrow 1}R_{i1\uparrow} - \sigma_{\downarrow 1}R_{i1\downarrow})\frac{q_u}{\sigma_{F1}}\}V_2^+ + \{\frac{1}{2}\alpha_{F1} - (\sigma_{\uparrow 1}R_{i1\uparrow} - \sigma_{\downarrow 1}R_{i1\downarrow})\frac{u_d}{\sigma_{F1}}\}V_2^-\eta_d$$

$$+(\sigma_{\uparrow 1}R_{i1\uparrow} + \sigma_{\downarrow 1}R_{i1\downarrow})\frac{j_e}{2\sigma_{F1}},$$

$$V_2^+ = \frac{\{-(Q_d+1)(R_{F1}\alpha_{F1} - R_{i1\uparrow} + R_{i1\downarrow}) + (U_d-1)\eta_d(R_{F3}\alpha_{F3} - R_{i3\uparrow} + R_{i3\downarrow})\}}{(U_u-1)(U_d-1)\eta_u\eta_d - (Q_d+1)(Q_u+1)}\frac{j_e}{2}$$

$$-\frac{\eta_d R_{F3}(U_d-1)(\sigma_{\uparrow 3}S_\uparrow - \sigma_{\downarrow 3}S_\downarrow)}{(U_u-1)(U_d-1)\eta_u\eta_d - (Q_d+1)(Q_u+1)}\frac{\nabla T_L}{2},$$

$$V_2^- = \frac{\{-(U_u-1)\eta_u(R_{F1}\alpha_{F1} - R_{i1\uparrow} + R_{i1\downarrow}) + (Q_u+1)(R_{F3}\alpha_{F3} - R_{i3\uparrow} + R_{i3\downarrow})\}}{(U_u-1)(U_d-1)\eta_u\eta_d - (Q_d+1)(Q_u+1)}\frac{j_e}{2}$$

$$-\frac{R_{F3}(Q_u+1)(\sigma_{\uparrow 3}S_\uparrow - \sigma_{\downarrow 3}S_\downarrow)}{(U_u-1)(U_d-1)\eta_u\eta_d - (Q_d+1)(Q_u+1)}\frac{\nabla T_L}{2}, \quad (9)$$

where $q_{d(u)} = (\frac{R_N + R_{d(u)}}{R_N R_{d(u)}})$, $u_{d(u)} = (\frac{R_N - R_{d(u)}}{R_N R_{d(u)}})$, $Q_{d(u)} = (R_F + R_{i3(i1)\uparrow} + R_{i3(i1)\downarrow})q_{d(u)}$,

$U_{d(u)} = (R_F + R_{i1(i3)\uparrow} + R_{i1(i3)\downarrow})u_{d(u)}$, $\eta_{N(d,u)} = \exp(-L/\lambda_{N(d,u)})$, $\alpha_{F1(3)} = \frac{\sigma_{\uparrow 1(3)} - \sigma_{\downarrow 1(3)}}{\sigma_{\uparrow 1(3)} + \sigma_{\downarrow 1(3)}}$ (the

spin polarization of the conductivity in FM1(3)). Here, the materials of FM1 and FM 3 are the same, and $\alpha_{F1} = \alpha_{F3}$. Under the anti-parallel configuration due to the magnetization reversal of FM3, we replace $\alpha_{F3}$, $\sigma_{\uparrow 3}$ and $R_{i3\uparrow}$ with $-\alpha_{F3}$, $\sigma_{\downarrow 3}$ and $R_{i3\downarrow}$ of FM3, respectively. The sum of the electric spin signals and the spincaloritronic signal from a non-degenerate Si spin valve in the 3T-MR scheme is quantified as the difference in $D$ between parallel and anti-parallel magnetic configurations, written as:

$$V_S = D^{parallel} - D^{antiparallel} = \left[\frac{\eta_d(\frac{1}{R_d}+\frac{1}{R_u})\{\frac{1}{2}\alpha_{F1}(R_{F1}+R_{i1}) - \frac{(\sigma_{\uparrow 1}R_{i1\uparrow} - \sigma_{\downarrow 1}R_{i1\downarrow})}{\sigma_{F1}}\}}{(U_u-1)(U_d-1)\eta_u\eta_d - (Q_d+1)(Q_u+1)}\right]$$

$$\times \{(R_{F3}\alpha_{F3} - R_{i3\uparrow} + R_{i3\downarrow})j_e - R_{F3}(\sigma_{\uparrow 3}S_{\uparrow 3} - \sigma_{\downarrow 3}S_{\downarrow 3})\nabla T_L\}, \quad (10)$$

where the second term shows the magnitude of the spincaloritronic signal.

**B. Electrical spin injection**

Electrical spin injection into non-degenerate Si is achieved by using a 3-terminal MRF scheme. The DC electric current of 3.0 mA without AC electric current was applied to the same sample discussed in the main text. A spin accumulation voltage was measured with sweeping an in-plain (parallel to the *y*-axis) magnetic field as shown in Fig. S1a (shown in the next page). The experimental result is shown in Fig. S1b, and the clear spin signal was observed. To corroborate successful spin transport in the Si, the Hanle measurement was conducted at the dc current of 3.0 mA, where the magnetic field perpendicular to the plain (parallel to the *z*-axis) was swept (see Fig. S2a in the next page). The clear Hanle signal was observed as shown in Fig. S2b, and the result was nicely reproduced by the following equation,

$$(1+\omega^2\tau'^2)^{-0.25}\exp\{\frac{L}{2\lambda_N}v\tau' - \frac{L}{\lambda_N}\sqrt{\frac{\sqrt{1+\omega^2\tau'^2}+1}{2}}\{\cos(\frac{\sqrt{(\arctan(\omega\tau'))^2}}{2}) + \frac{L}{\lambda_N}\sqrt{\frac{\sqrt{1+\omega^2\tau'^2}-1}{2}}\},$$

where $D$ is the spin diffusion constant, $\tau$ is the spin lifetime, $\omega = g\mu_B B/\hbar$ is the Larmor frequency, $g$ is the $g$-factor for the electrons ($g = 2$ in this study), $\mu_B$ is the Bohr magneton, $\hbar$ is the Dirac constant, $v$ is the spin drift velocity, and $\tau' = \frac{v^2}{4D} + \frac{1}{\tau}$ is the modified spin lifetime by the spin drift. The spin lifetime, $\tau$, and the spin diffusion length, $\lambda$, in the Si were estimated to be 6.4 ns and 5.0 μm, respectively.

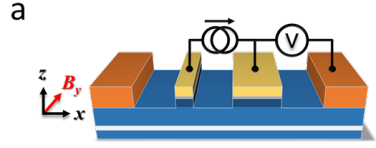
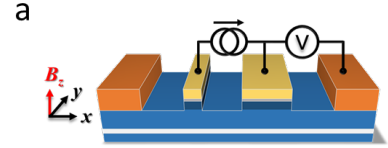
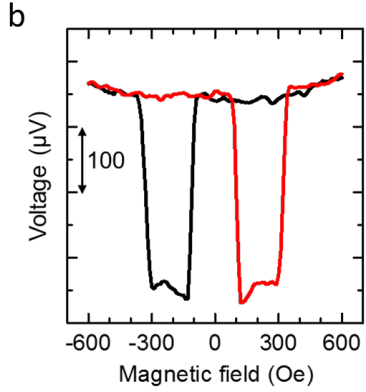
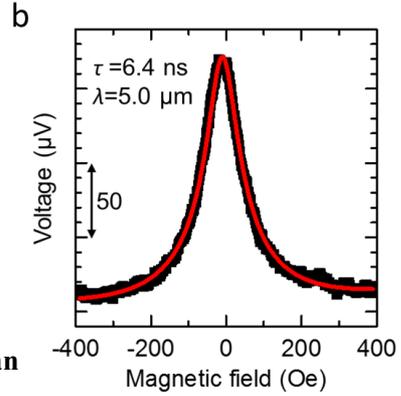

**C.** Detail of an estimation of spurious signals.

To eliminate spurious signals and estimate the magnitude of the spincaloritronic signals precisely, we carried out the following analyses. As described in the main text, $I(V)$ curve was fitted using $5^{th}$ order polynomial function. The result of the $I$-$V$ curve fitting (shown in Fig. 3a) is summarized in Table I. The coefficient of $V^5$ is small enough and we concluded that the $5^{th}$ order polynomial function is appropriate to reproduce the experimental results. Spin signal dependence as a function of applied DC current was also measured and fitted using $5^{th}$ order polynomial function. We injected previous coefficient from $I(V)$ in order to obtain the dependence of spin signal in function of applied voltage: $V_{DC\ spin\ signal} = \sum_{i=1}^{5} R_{s,i} (\sum_{k=1}^{5} G_k(V_{inj}))$. Because this signal includes spurious signals measured in the spincaloritronic experiment, we calculate the time dependence of this signal to evaluate the contribution of the spurious signal at $2f$ in our experiment. To do that we defined the time dependence of applied voltage as: $V_{inj}(t) = V_{dc} + V_0 \sin(2\pi f t)$, the DC voltage was 3.0 V and the AC voltage was 1.0 V at 17 Hz.

Thus, the time-dependent spectrum and the Fourier transformation spectra of $V_{DC\ spin\ signal}$ are calculated as shown in Fig. S3. The 1$f$ signal is obtained to be 106 μV, which is consistent with experimental result (ca. 110 μV). On the contrary, the 2$f$ signal is 10.7 μV, inconsistent with the experimental result (ca. 19 μV). Because the calculated 2$f$ signal is only due to electrical spin injection or the spurious effects, not the spincaloritronic effect, the difference of the measured and the calculated 2$f$ signal is thus the spincaloritronic signal.

**Figure S3. (a)** Time-dependent spectrum of the electric spin signal. **(b)** Fourier transformation spectrum of the electric spin signal.

**Figure S4.** 1f spin signal at dc voltage of 3.0 V, ac voltage of 1.0 V and 17 Hz.

**Table I.** Coefficients for each order of the polynomial function.

**D. Comparison of the result using "Seebeck spin tunneling effect" in Si and our result**

A spincaloritronic effect in semiconductor, the "Seebeck spin tunneling" effect [S1], was reported by Le Breton et al. The measuring method was a so-called electrical 3-terminal (3T) method (different from 3T-MR used in this study), which was also used in electrical "spin accumulation" experiment by the same group [S2]. Le Breton et al. showed that linewidth and shapes of open-circuit Hanle-like voltage signals in the "Seebeck spin tunneling" and the "electrical 3T" experiments were the same, and they claimed that it confirmed the successful "spin accumulation" in the "Seebeck spin tunneling" experiments. However, one should note that the linewidth of the open-circuit Hanle-like voltages from n-type and p-type Si in both the

"Seebeck spin tunneling" and the "electrical 3T" experiments was the same. The linewidth should be governed by the spin lifetime in the channel (thus, the lifetime can be estimated by the Hanle-effect-related Lorenzian function fitting). It is well known that in silicon the spin lifetime of electrons in the conduction band is orders of magnitude larger than that of holes in the valence band (since states in the valence band are not pure spin states, and almost any momentum scattering event leads to the spin flip) [S3, S4]. In fact, the spin lifetime of the accumulated spins in n-Si (the doping concentration $1.8 \times 10^{19}$ cm$^{-3}$) using the 3-terminal method was reported to be 142 ps at RT [S2], and the spin lifetime of transported spins in n-Si (the doping concentration $5 \times 10^{19}$ cm$^{-3}$) was estimated to be 1.3 ns by using a non-local 4-terminal method [S3]. The doping concentrations of both Si were almost the same, and furthermore, the Elliot-Yafet spin relaxation takes place in Si, i.e. Si with a lower doping concentration should exhibit longer spin lifetime. In contrast to 3T method, in non-local 4T method the actual spin transport is realized and measured, thus, it is one of the most precise methods to measure spin lifetime. Large discrepancy in the spin lifetime between two methods evidenced the non-spin accumulation origin of the signals in the 3T method. Regarding the reliability of the 3T signals, see also the paper by Aoki et al. [S5]. Hence, the similarity of the data in the two studies of refs. [S1] and [S2] showed that the spin lifetime cannot be extracted using the Hanle-effect-related Lorentzian function in 3T experiments, while the origin of this 3T open circuit voltage signal remained unclear [S5, S6]. In follow-up experiments, other groups using the same 3T scheme obtained similar magnetoresistance not only from semiconductor/ferromagnet tunnel junction but also from nonmagnet/ferromagnet junctions, where the spin-orbit strength varies drastically between different nonmagnet metals [S7-S12].

All of the observed 3T open-circuit Hanle-like voltage signals had amplitude orders of magnitude larger than a spin accumulation signal, and their width was roughly the same: in contradiction to a spin accumulation signal whose width reflects the spin lifetime. This discrepancy triggered theoretical efforts from multiple groups aimed to understand the origin of a 3T open circuit voltage signal [S9, S12]. They clarified that such signal arises from spins captured by impurity and trap levels in oxide tunneling barriers and/or a modulation of tunneling resistance of the oxide barriers by a magnetic field, and non-related to spin accumulation. Thus, in our study, we for the first time demonstrated the spin accumulation and transport in a semiconductor channel (Si) using a spincaloritronic effect.